\documentstyle[preprint,aps,prb]{revtex}

\begin{document}
\draft
\title{Spin relaxation in asymmetrical heterostructures}

\author{N.S.~Averkiev,$^1$ L.E.~Golub\footnote{E-mail: golub@coherent.ioffe.rssi.ru},$^{1,2}$ and M.~Willander$^2$} 
\address{$^1$A.F.~Ioffe Physico-Technical Institute, Russian Academy  of Sciences, 194021 St.~Petersburg, Russia}
\address{$^2$Physical Electronics and Photonics, Department of Physics,
Chalmers University of Technology and G\"{o}teborg University, S-412~96, G\"{o}teborg, Sweden}
\date{\today}
\maketitle

\begin{abstract}
Electron spin relaxation caused by the D'yakonov-Perel' mechanism is investigated theoretically in asymmetrical A$_3$B$_5$ heterostructures. The total spin relaxation anisotropy is demonstrated for a wide range of structure parameters and temperatures. The spin relaxation rates dependences are derived for GaAs-based heterojunction and triangular quantum well. The calculations show a few orders of magnitude difference in spin relaxation times.
\end{abstract}

\section{Introduction}
The degrees of freedom of spin have a great deal of attention throughout the development of semiconductor physics. Recently the spin properties of carriers have been investigated intensely in low-dimensional semiconductor structures. In electronics, much interest in spin  has also appeared due to  recent proposals to construct spin transistors and spin computers based on heterostructures.~\cite{Nitta,Kane}

Spin-orbit interaction, which determines a spin behavior,  is much more complex in semiconductor heterostructures than in bulk systems. The bulk spin-orbit terms take a more interesting form in two-dimensional systems, and, in addition, new terms appear which are absent in bulk.

In Ref.~\onlinecite{PRB} we considered electron spin dynamics in asymmetrical heterostructures. Giant anisotropy of spin relaxation times caused by the interference of different spin-orbit terms has been revealed. In this work, we calculate spin relaxation rates in real asymmetrical structures. A heterojunction and a triangular quantum well are considered in detail. The effect of heteropotential asymmetry on spin relaxation is investigated in a wide range of electron concentrations and temperatures. We show that giant spin relaxation anisotropy governed by external parameters, opening possibility for applications in spin engineering.

\section{Theory}

Consider a system with spin-orbit interaction described by the Hamiltonian $H_{\rm SO} ({\bbox k})$, where ${\bbox k}$ is a wavevector. $H_{\rm SO} ({\bbox k})$ is equivalent to a Zeeman term with an effective magnetic field dependent on ${\bbox k}$. In the presence of scattering, the wavevector changes and, hence, the effective magnetic field changes too. Therefore for frequent scattering, the electrons are subjected to a chaotically changing magnetic field. The spin dynamics in such a system has a diffusion character, which leads to loss of the specific spin orientation. This is called the D'yakonov-Perel'  spin relaxation mechanism,~\cite{DP} and it is the main spin relaxation mechanism in many A$_3$B$_5$ bulk semiconductors and heterostructures.

For a two-dimensional system with any  $H_{\rm SO} ({\bbox k})$ (where ${\bbox k}$ lies in the plane of the heterostructure), one can show similarly to Ref.~\onlinecite{PRB}, (see also Refs.~\onlinecite{DP,Pikus/Titkov,DK}), that the spin dynamics of electrons in the presence of elastic scattering is described by the following equations
\begin{equation}
\label{S}
\dot{S}_i(t) = - {1 \over 2 \hbar^2}\sum\limits_{n=-\infty}^\infty \frac{ \int\limits_0^\infty d \varepsilon  \: (F_+ - F_-) \: \tau_n \: Tr \left\{[H_{-n},[H_n, \sigma_j]] \: \sigma_i \right\}}{ \int\limits_0^\infty d \varepsilon \: (F_+ - F_-)}\: S_j(t) \:. \end{equation}
Note that this is true only for times longer than the momentum relaxation time but shorter than the spin relaxation times.
In~(\ref{S}), $S_i$ are the spin density components ($i=x,y,z$), the integration is performed over the energy $\varepsilon = \hbar^2 k^2 / 2 m$, where $m$ is the electron effective mass, $F_\pm (\varepsilon)$ are distribution functions of electrons with the spin projection equal to $\pm 1/2$, $\sigma_i$  are the Pauli matrices. $H_n$ are the harmonics of the spin-orbit Hamiltonian:
\begin{equation}
\label{H/n}
H_n = \oint \: {d \varphi_{\bf k} \over 2 \pi} \: H_{\rm SO}({\bbox k}) \exp (-i n \varphi_{\bf k} ) \:, 
\end{equation}
where $\varphi_{\bf k}$ is the angular coordinate of ${\bbox k}$, and the scattering times are
\begin{equation}
\label{tau/n}
{1 \over \tau_n} = \oint \: d \theta \: W(\varepsilon, \theta) (1 - \cos n \theta ) \:, \end{equation}
where $W(\varepsilon, \theta)$ is the probability of elastic scattering by an angle $\theta$ for an electron with energy $\varepsilon$.

Equation~(\ref{S}) is valid for two-dimensional electrons with any spin-orbit interaction $H_{\rm SO}({\bbox k})$. Now we consider an asymmetrical zinc-blende heterostructure. There are two contributions to $H_{\rm SO}({\bbox k})$. The first, the so-called bulk inversion asymmetry (BIA) term, is due to lack of inversion symmetry in the bulk material from which the heterostructure is made. To calculate this term, one has to average the corresponding bulk expression over the size-quantized motion.~\cite{DK} We investigate the heterostructure with the growth direction [001] coinciding with the $z$-axis and assume that only the first electron subband is populated. The BIA-term has the form
\begin{equation}
\label{H1}
H_{\rm BIA}({\bbox k}) = \gamma \: [ \sigma_x k_x (k_y^2 - \langle k_z^2 \rangle) + 
\sigma_y k_y (\langle k_z^2 \rangle - k_x^2)] \:,
\end{equation}
where we choose $x$- and $y$-directions to be along the principal axes in the plane of the heterostructure. Here $\langle k_z^2 \rangle$ is the square of the operator $(- i \partial / \partial z)$ averaged over the ground state, and  $\gamma$ is the bulk spin-orbit interaction constant.   It is seen that $H_{\rm BIA}$ contains terms both linear and cubic in $k$.

In asymmetrical heterostructures, there is an additional contribution to the spin-orbit Hamiltonian which is absent in the bulk. It is caused by structure inversion asymmetry (SIA) and can be written as~\cite{OhkawaUemura,Vas'ko,BychkovRashba}
\begin{equation}
\label{H2}
H_{\rm SIA}({\bbox k})  = \alpha \: (\sigma_x k_y - \sigma_y k_x) \:,
\end{equation}
where $\alpha$ is proportional to the average electric field, $E$, acting on an electron:
\begin{equation}
\label{alpha}
\alpha = \alpha_0 e  E \:.
\end{equation}
Here $e$ is the elementary charge and $\alpha_0$ is a second spin-orbit constant determined by both bulk spin-orbit interaction parameters and properties of  heterointerfaces.

$H_{\rm SIA}$ also contains  terms linear in $k$. From the~(\ref{S}) it follows that the harmonics with the same $n$ are coupled in the spin dynamics equations. This leads to the interference of linear in wavevector BIA- and SIA-terms in spin relaxation.~\cite{PRB}

For $H_{\rm SO} = H_{\rm BIA} + H_{\rm SIA}$, the system has the $C_{2v}$-symmetry. Therefore equations~(\ref{S}) may be rewritten as follows:
\begin{equation}
\label{S1}
\dot{S}_z = - {S_z \over \tau_z} \:,\hspace{2cm} \dot{S}_x \pm \dot{S}_y = - {S_x \pm S_y \over \tau_\pm} \:.
\end{equation}
The times $\tau_z$, $\tau_+$ and $\tau_-$  are the relaxation times of the spin parallel to the axes [001], [110] and [1$\bar{1}$0], respectively.

If both spin subsystems come to equilibrium before the start of the spin relaxation, then
\begin{equation}
\label{F}
F_\pm (\varepsilon) = F_0 (\mu_\pm -  \varepsilon)\:,
\end{equation}
where $F_0$ is the Fermi-Dirac distribution function and $\mu_\pm$ are the chemical potentials of electron spin subsystems. If the spin splitting is small, i.e.
$$
|\mu_+ - \mu_-| \ll |\mu_+|, |\mu_-|
$$
then the expressions for the spin relaxation rates $1/ \tau_i$ ($i = z, +, -$) will have the form
\begin{equation}
\label{tau}
{1 \over \tau_i} = \frac{ \int\limits_0^\infty d \varepsilon  \: (\partial F_0 / \partial \varepsilon) \: \Gamma_i (k)}{ \int\limits_0^\infty d \varepsilon \: (\partial F_0 / \partial \varepsilon)}\:, \end{equation}
where
\begin{eqnarray}
\label{Gamma}
\Gamma_z (k) = {4 \tau_1 \over \hbar^2}\left[  (\gamma^2 \langle k_z^2 \rangle^2 +\alpha^2) \: k^2 - {1 \over 2} \gamma^2 \langle k_z^2 \rangle k^4 + {1 + \tau_3/\tau_1 \over 16} \gamma^2 k^6 \right]\:, \\
\Gamma_\pm (k) = {2 \tau_1 \over \hbar^2}\left[  (\pm \alpha - \gamma \langle k_z^2 \rangle)^2 \: k^2 + {1 \over 2} \gamma (\pm \alpha - \gamma \langle k_z^2 \rangle) k^4 + {1 + \tau_3/\tau_1 \over 16} \gamma^2 k^6 \right]\:. \nonumber
\end{eqnarray}

Eqs.~(\ref{tau},\ref{Gamma}) are valid for any electron energy distribution. If the electron gas is degenerate, then the spin relaxation times are given by
\begin{equation}
\label{tauFermi}
{1 \over \tau_i} = \Gamma_i (k_{\rm F})\:, 
\end{equation}
where $k_{\rm F}$ is the Fermi wavevector determined by the total two-dimensional electron concentration, $N$:
\begin{equation}
\label{kF}
k_{\rm F} = \sqrt{2 \pi N}\:.
\end{equation}
In this case, the scattering time $\tau_1$ in Eqs.~(\ref{Gamma}) coincides with the transport relaxation time, $\tau_{\rm tr}$, which can be determined from the electron mobility.

For non-degenerate electrons, the spin relaxation times are determined, in particular, by the energy dependences of scattering times $\tau_1$ and $\tau_3$. If $\tau_1$, $\tau_3$ $\sim$ $\varepsilon^\nu$, then $\tau_3/\tau_1 =$~const and
\begin{eqnarray}
\label{tauBoltzmann}
{1 \over \tau_z} = {4 \tau_{\rm tr} \over \hbar^2}\left[  (\gamma^2 \langle k_z^2 \rangle^2 +\alpha^2) \: 
{2 m k_{\rm B}T \over \hbar^2} 
- {\nu+2 \over 2} \gamma^2 \langle k_z^2 \rangle \: \left({2 m k_{\rm B}T \over \hbar^2} \right)^2 \right. \nonumber\\
\left. + (\nu+2)(\nu+3) \: {1 + \tau_3/\tau_1 \over 16} \gamma^2 \: \left({2 m k_{\rm B}T \over \hbar^2} \right)^3 \right]\:, \\
{1 \over \tau_\pm} = {2 \tau_{\rm tr} \over \hbar^2}\left[  (\pm \alpha - \gamma \langle k_z^2 \rangle)^2  \: 
{2 m k_{\rm B}T \over \hbar^2}
+ {\nu+2 \over 2} \gamma (\pm \alpha - \gamma \langle k_z^2 \rangle) \: \left({2 m k_{\rm B}T \over \hbar^2} \right)^2  \right.
\nonumber \\
\left. +  (\nu+2)(\nu+3) \: {1 + \tau_3/\tau_1 \over 16} \gamma^2 \: \left({2 m k_{\rm B}T \over \hbar^2} \right)^3 \right]\:. \nonumber
\end{eqnarray}
Here $T$ is the electron temperature and $k_{\rm B}$ is the Boltzmann constant. In the particular case of short-range scattering, $\nu = 0$, and $\tau_1 = \tau_3$  are equal to $\tau_{\rm tr}$, which does not depend on temperature.

Spin relaxation times are very sensitive to the relationship between two spin-orbit interaction strengths, $\gamma \langle k_z^2 \rangle$ and $\alpha$. From Eqs.~(\ref{tauFermi}),~(\ref{tauBoltzmann}) it follows that at low concentration or temperature, $1/\tau_z$,  $1/\tau_-$ and  $1/\tau_+$ are determined by the sum of the squares, by the square of the sum, and by the square of the difference of $\gamma \langle k_z^2 \rangle$ and $\alpha$, respectively. This can lead to a large difference in the three times, i.e. to a total spin relaxation anisotropy, if $\gamma \langle k_z^2 \rangle$ and $\alpha$ are close in magnitude.

In real A$_3$B$_5$ systems, the relationship between $H_{\rm BIA}$ and $H_{\rm SIA}$ may be different. $H_{\rm BIA}$ or $H_{\rm SIA}$ may be dominant,~\cite{BIAdominant,SIAdominant}or they may be comparable.~\cite{Knap}

The value of $\langle k_z^2 \rangle$ depends on the shape of the heteropotential and will be calculated for given asymmetrical heterostructures below. The constant $\gamma$ is known for GaAs from optical orientation experiments.~\cite{Pikus/Titkov} The correct theoretical expressions for $\gamma$ and $\alpha_0$ were derived using the three-band ${\bbox k} \cdot {\bbox p}$ model.~\cite{Knap} The heterointerfaces give a contribution to $\alpha_0$ in addition to that of the bulk.~\cite{Silvanew} 
At large wavevectors, $\alpha_0$ starts to depend on $k$.~\cite{Roessler1,Roessler2} Here we assume concentrations and temperatures sufficiently low allowing us to ignore this effect.

The spin relaxation rates for two types of asymmetrical structures | for a heterojunction and a triangular quantum well | are calculated below. The scattering assumed to be short-range ($\nu = 0$, $\tau_3 = \tau_1 = \tau_{\rm tr}$). All parameters are chosen to be correspond to GaAs/AlAs heterostructure: $\gamma = 27$~eV$\cdot$\AA$^3$, $m = 0.067 m_0$, where $m_0$ is the free electron mass, $\alpha_0 = 5.33$~\AA$^2$. The time $\tau_{\rm tr}$ is taken equal to 0.1~ps and assumed to be independent of electron concentration. 

\section{Spin relaxation in a heterojunction}

In a heterojunction, the size of spin-orbit interaction is governed by the two-dimensional carrier concentration, $N$. $\langle k_z^2\rangle$ may be estimated as follows:~\cite{AFS}
\begin{equation}
\label{kz2}
\langle k_z^2\rangle = {1 \over 4}\left( {16.5 \pi N e^2 m \over \kappa \hbar^2} \right)^{2/3}\:,
\end{equation}
where $\kappa$ is the dielectric constant. The mean electric field acting on an electron may be taken to be equal to half of the maximum field in the junction:
\begin{equation}
\label{E}
E = {2 \pi N e \over \kappa}\:.
\end{equation}

Fig.~\ref{tau_hetero} shows the concentration dependence of the reciprocal spin relaxation times for degenerate electrons in  GaAs/AlAs heterojunction ($\kappa =12.55$). The inset shows the spin-orbit interaction strengths, $\gamma \langle k_z^2 \rangle$ and $\alpha$, and the absolute value of their difference, as functions of electron concentration. 

One can see the total spin relaxation anisotropy over a wide range of concentrations. At small $N$, $1/\tau_+$ is less than $1/\tau_-$, whereas at large concentration $1/\tau_+$ is greater than $1/\tau_-$. This is due to the factor which is multiplied by $k^2$ in~(\ref{Gamma}) being larger for $1/\tau_-$ than for $1/\tau_+$ and the reverse being true for the factors which are multiplied by $k^4$. Therefore at a certain concentration, the times $\tau_+$ and $\tau_-$  will be equal. From Eqs.~(\ref{Gamma}),~(\ref{tauFermi}) follows that this takes place when 
\begin{equation}
\label{isotr/degenerate}
k_{\rm F}^2 = 4 \langle k_z^2 \rangle \:,
\end{equation}
which is fulfilled at $N = 1.1 \times 10^{13}$~cm$^{-2}$ as illustrated in Fig.~\ref{tau_hetero}. At larger concentrations, spin relaxation is again totally anisotropic.

Despite the values $\gamma \langle k_z^2 \rangle$ and $\alpha$ are close in magnitude over a wide range of concentrations (see the inset in Fig.~\ref{tau_hetero}), all three spin relaxation rates depend on $N$ monotonically. This happens because, as the concentration  increases, $k_{\rm F}$ increases as well, and the terms in $H_{\rm SO}$ which are cubic in the wavevector become important. The growth of these terms with $N$ dominates the change of the function  $(\alpha - \gamma \langle k_z^2 \rangle)^2$ in~(\ref{Gamma}), hence the monotonous concentration dependence of $1/\tau_+$ occurs.

The situation changes in the case of Boltzmann gas. For non-degenerate electrons, the mean value of wavevector and the concentration are independent. For temperatures up to 300~K, the characteristic $k^2 \sim 2 m k_{\rm B} T / \hbar^2$ is much less than $\langle k_z^2 \rangle$, and the spin relaxation rates are determined by the first terms in~(\ref{tauBoltzmann}). As a result, all three spin relaxation times are different up to 300~K at a given concentration. The corresponding calculations are presented in Fig.~\ref{tau_T_hetero_vsT}.

The times $\tau_+$ and $\tau_-$ are equal to each other at a certain temperature only. According to~(\ref{tauBoltzmann}), the corresponding condition is
\begin{equation}
\label{isotr/Boltzmann}
T = {\hbar^2 \langle k_z^2 \rangle \over m k_{\rm B} (1 + \nu/2)} \:.
\end{equation}
Using the GaAs-parameters and $\nu = 0$ in~(\ref{kz2}) , it can be seen that~(\ref{isotr/Boltzmann})  is satisfied for $T \approx 100$~K for $N=10^{11}$~cm~$^{-2}$ and at $T \approx 290$~K for $N=5 \times 10^{11}$~cm~$^{-2}$, which agrees very well with Fig.~\ref{tau_T_hetero_vsT}.

At a fixed temperature, the spin relaxation rates are governed by the electron concentration. According to Eqs.~(\ref{tauBoltzmann}), the dependences of $1/\tau_i$ on $N$ are similar to the curves in the inset in Fig.~\ref{tau_hetero}. In particular, from Eqs.~(\ref{tauBoltzmann}) follows that $1/\tau_z$ and $1/\tau_-$ have to be close in magnitude, and they both greatly exceed $1/\tau_+$. In addition, $1/\tau_+$ depends on concentration non-monotonically. This is confirmed completely by the results presented in Fig.~\ref{tau_T_hetero}. One can see that $1/\tau_+ \ll 1/\tau_z \approx 1/\tau_-$, and the rate $1/\tau_+$ has a minimum when plotted as a function of concentration. This minimum is at $N=1.4 \times 10^{13}$~cm$^{-2}$, when the terms in $H_{\rm SO}$ linear in the wavevector cancel each other out. The corresponding condition is
\begin{equation}
\label{cancellation}
\gamma \langle k_z^2\rangle = \alpha \:.
\end{equation}
At this concentration, the spin relaxation time $\tau_+$ is very large but remains finite due to the terms cubic in $k$. Therefore the difference in the spin relaxation times is more pronounced at low temperature. At high $T$, the terms in $H_{\rm SO}$ cubic in the wavevector become significant, and the minimum in $1/\tau_+$ disappears. However $1/\tau_+$ is still much less than $1/\tau_-$, i.e. huge spin relaxation anisotropy occurs in the plane of  the heterojunction even at room temperature.

\section{Spin relaxation in a triangular quantum well}

In this Section, we investigate spin relaxation in the following asymmetrical system. We consider a structure with infinitely-high barrier at $z < 0$ and constant electric field, $E$, at $z > 0$. 

In the framework of this model, 
\begin{equation}
\label{kz2_triangular}
\langle k_z^2\rangle = a \: \left( {2 m e E \over \hbar^2} \right)^{2/3} \:,
\end{equation}
where
\begin{equation}
a =  \frac{\int\limits_0^\infty dx  \: [Ai'(x-\beta)]^2}{\int\limits_0^\infty dx  \: [Ai(x-\beta)]^2} \approx 0.78 \: .
\end{equation}
Here $(-\beta)$ is the first root of the Airy function:
\[
Ai(-\beta)=0\:, \hspace{2cm} \beta \approx 2.338 \: .
\]
The value of $\alpha$ is given by~(\ref{alpha}).

In Fig.~\ref{tau_triang} the spin relaxation rates  are plotted for the triangular GaAs quantum well at different electric fields. It can be seen that the total spin relaxation anisotropy occurs for both degenerate and Boltzmann gases in wide ranges of concentrations and temperatures. The times $\tau_+$ and $\tau_-$ coincide only at a specific concentration or temperature. For degenerate electrons, according to Eq.~(\ref{isotr/degenerate}), the corresponding curves have an intersection at $N \approx 3.4 \times 10^{12}$~cm~$^{-2}$ for $E = 10^5$~V/cm and at $N \approx 7 \times 10^{12}$~cm~$^{-2}$ for $E = 3 \times10^5$~V/cm in agreement with Fig.~\ref{tau_triang}a. For a Boltzmann gas, the intersection of $\tau_+$ and $\tau_-$ occurs at $T \approx 150$~K for $E = 10^4$~V/cm and at $T \approx 240$~K for $E = 2 \times 10^4$~V/cm, according to~(\ref{isotr/Boltzmann}). This is also confirmed by Fig.~\ref{tau_triang}b.

The behavior of the reciprocal spin relaxation times in electric field is presented in Fig.~\ref{tau_triang_vsE} for both degenerate and Boltzmann electron gas. The dependences of  $\gamma \langle k_z^2\rangle$ and $\alpha$ on electric field are similar to the inset in Fig.~\ref{tau_hetero}: their values are close to each other in magnitude, so the difference between them is very small. This leads to the minimum in the dependence of $1/\tau_+$ on $E$. The cancellation condition~(\ref{cancellation}) is fulfilled at $E \approx 1.9 \times 10^6$~V/cm. The electric field of this size can be created in heterostructures containing a gate, allowing experimental observation of non-monotonic spin relaxation rate dependence shown in  Fig.~\ref{tau_triang_vsE}.

\section{Conclusion}
It has been shown~\cite{LMR,Silvaoldest,Jusserand} that inclusion of both BIA and SIA terms~(\ref{H1}) and~(\ref{H2}) into $H_{\rm SO}$ leads to the conduction band spin-splitting anisotropy in ${\bbox k}$-space in A$_3$B$_5$ semiconductor heterojunctions. However the spin relaxation analysis that has been performed~\cite{Silvaoldest} ignored this effect.

The authors of Ref.~\onlinecite{Pikus/theor} showed that the BIA and SIA terms interfere in weak localization but are additive in spin relaxation. In this paper, we prove that the terms in $H_{\rm SO}$ linear in the wavevector cancel each other in spin relaxation as well.

In a recent experiment,~\cite{Japan110} the spin relaxation anisotropy was observed for non-commonly used (110) GaAs quantum wells. In this experiment, the spin relaxation in the growth direction was suppressed because of the ``built-in'' anisotropy of the sample due to presence of  heterointerfaces. In the present paper, we predict spin relaxation suppression in the plane of a heterostructure. Moreover, all three spin relaxation times are different in our case, and this effect takes place in ordinary (001) heterostructures.

To observe the predicted spin relaxation anisotropy, one can perform time-resolved measurements similar to Ref.~\onlinecite{Japan110}. In steady-state experiments, spin relaxation can be investigated via the Hanle effect. To obtain the spin relaxation times, one has to take into account of the fact that the Land\'{e} $g$-factor has not only diagonal in-plane components ($g_{xx}$) but also off-diagonal ones ($g_{xy}$) in asymmetrical heterostructures.~\cite{KK} The degree of polarization of the photoluminescence in a magnetic field ${\bbox B} \perp z$ is described by the following expression
\begin{equation}
\label{Hanle}
P ({\bbox B}) = \frac{P(0)}{1 + [\mu_B \: (g_{xx} \pm g_{xy}) \: B]^2 \: \tau_z \: \tau_\mp} \:,
\end{equation}
where the upper and lower signs correspond to the experimental geometry ${\bbox B} || [110]$ and ${\bbox B} || [1\bar{1}0]$, respectively ($\mu_B$ is the Bohr magneton).

We show that the terms in the spin-orbit Hamiltonian which are linear in the wavevector interfere which leads to the huge anisotropy of the spin relaxation times. At a high concentration or temperature, this effect starts to disappear due to domination of the terms in $H_{\rm SO}$ cubic in $k$ which are present only in $H_{\rm BIA}$. However the higher order terms in $H_{\rm SIA}$ are not forbidden by symmetry either. These terms can also interfere with these in $H_{\rm BIA}$, and cause additional non-monotonic peculiarities in spin relaxation times dependences on the structure parameters.

In conclusion, we calculated the spin relaxation times for an A$_3$B$_5$ heterojunction and triangular quantum well. The observance of total spin relaxation anisotropy is predicted in a wide range of structure parameters and temperatures.

\section*{Acknowledgements} 

We thank J.~Vincent for critical reading of the manuscript.
This work was supported by the Russian Foundation of Basic Research, projects 00-02-17011 and 00-02-16894 and the Russian State Programme ``Physics of Solid State Nanostructures''.

\begin{figure}
\caption{
\label{tau_hetero}
The concentration dependences of the reciprocal spin relaxation times, $1/\tau_z$ (solid line), $1/\tau_-$ (dashed line) and $1/\tau_+$ (dotted line), for GaAs/AlAs heterostructure at zero temperature. The parameters are given in the text. The inset shows the spin-orbit interaction strengths, $\gamma \langle k_z^2 \rangle$ (solid line), $\alpha$ (dashed), and $|\gamma \langle k_z^2 \rangle - \alpha|$ (dotted),  in eV$\cdot$\AA, as functions of electron concentration, $N / (10^{12}$~cm$^{-2}$).
}
\end{figure}

\begin{figure}
\caption{
\label{tau_T_hetero_vsT}
The temperature dependences of spin relaxation rates, $1/\tau_-$ (solid line), $1/\tau_z$ (dashed line) and $1/\tau_+$ (dotted line),  for GaAs/AlAs heterostructure at different electron concentrations.
}
\end{figure}

\begin{figure}
\caption{
\label{tau_T_hetero}
The concentration dependences of the reciprocal spin relaxation times, $1/\tau_+$ (solid line), $1/\tau_z$ (dashed line) and $1/\tau_-$ (dotted line), for Boltzmann electron gas in GaAs/AlAs heterostructure at different temperatures. ({\em 1}) $T=30$~K, ({\em  2}) $T=77$~K, ({\em 3}) $T=150$~K, ({\em 4}) $T=300$~K.
}
\end{figure}

\begin{figure}
\caption{
\label{tau_triang}
The spin relaxation rates, $1/\tau_z$ (solid line), $1/\tau_-$ (dashed line) and $1/\tau_+$ (dotted line), in the triangular GaAs quantum well at different electric field. $a$ | degenerate electron gas, $b$ | Boltzmann gas.
}
\end{figure}

\begin{figure}
\caption{
\label{tau_triang_vsE}
The spin relaxation rates, $1/\tau_+$ (solid line), $1/\tau_z$ (dashed line) and $1/\tau_-$ (dotted line), in the triangular GaAs quantum well as functions of  the electric field. 
$a$ | degenerate electrons, ({\em 1}) $N=10^{11}$~cm$^{-2}$, ({\em 2})  $N=3\times10^{11}$~cm$^{-2}$, ({\em 3}) $N=5\times10^{11}$~cm$^{-2}$, ({\em 4}) $N=10^{12}$~cm$^{-2}$. $b$ | Boltzmann gas, ({\em 1}) $T=30$~K, ({\em 2}) $T=77$~K, ({\em 3}) $T=150$~K, ({\em 4}) $T=300$~K.
}
\end{figure}
\end{document}